\documentclass[a4paper,11pt]{article}
\usepackage{jcappub} 

\usepackage{Style} 
\usepackage{slashed}
\usepackage[dvipsnames]{xcolor}
\usepackage[caption=false]{subfig}
\usepackage{subfig}

\arxivnumber{2506.23244 } 
\title{Supersymmetric Hybrid Inflation with K\"{a}hler-Induced $\mathbf{R}$-Symmetry Breaking}

\author[a]{Muhammad Nadeem Ahmad 
}
\affiliation[a]{Department of Physics, Quaid-i-Azam University, \\
Islamabad 45320, Pakistan}

\author[b, 1]{and Mansoor Ur Rehman
\note{Corresponding author.}}
\affiliation[b]{Department of Physics, Faculty of Science, Islamic University of Madinah, \\
Madinah 42351, Saudi Arabia}
\emailAdd{mansoor@qau.edu.pk}

\abstract{We explore the role of explicit nonrenormalizable $R$-symmetry breaking interactions in the context of supersymmetric hybrid inflation. In particular, we focus on scenarios where such breaking arises predominantly from the K\"{a}hler potential, while the renormalizable terms in both the superpotential and K\"{a}hler potential preserve $R$-symmetry. Incorporating radiative corrections, soft SUSY-breaking contributions, and supergravity effects, we construct a consistent and predictive inflationary framework. Notably, the presence of $R$-symmetry violating terms at the nonrenormalizable level helps resolve the common issue of light waterfall fields in grand unified theories, rendering them sufficiently heavy without disturbing gauge coupling unification. Our numerical analysis demonstrates that these $R$-symmetry breaking contributions play a crucial role in bringing the scalar spectral index $n_s$ into excellent agreement with the recent cosmological observations, particularly the Data Release 6 of the Atacama Cosmology Telescope. The tensor-to-scalar ratio remains suppressed, with $r < 10^{-5}$, below the reach of current and near-future experiments. However, observable gravitational waves with $r \lesssim 0.03$ can be achieved by allowing moderate deviations in the parameter space associated with a non-minimal K\"{a}hler potential.}

\begin{document}
\maketitle
\flushbottom

\section{Introduction} \label{sec:introduction}

Supersymmetric (SUSY) hybrid inflation (SHI) \cite{Dvali:1994ms, Copeland:1994vg, Linde:1997sj, Buchmuller:2000zm, Senoguz:2003zw, Senoguz:2004vu} has been actively studied in the phenomenological framework of various models that differ in the choice of underlying gauge symmetry, mechanism of symmetry breaking, chronology of end of inflation versus symmetry breaking, or presence of additional local or global symmetries, etc. Initially utilized to explain the experimental inflationary data in the SUSY context \cite{Kyae:2005nv, Bastero-Gil:2006zpr, Rehman:2009nq, Civiletti:2014bca}, the SHI paradigm gradually evolved and was expanded to additionally investigate the phenomena like proton decay \cite{Rehman:2009yj, Lazarides:2020bgy, Mehmood:2020irm, Abid:2021jvn, Ijaz:2023cvc, Ahmed:2024iyd}, observable gravitational waves \cite{Shafi:2010jr, Rehman:2010wm, Rehman:2018nsn}, as well as consistency with modified swampland and trans-Planckian censorship conjectures \cite{Ahmed:2024rdd}, among others. 

Different variants of SHI that have been worked out and explored include smooth hybrid inflation \cite{Lazarides:1995vr, urRehman:2006hu, Rehman:2012gd, Rehman:2014rpa, Zubair:2024quc, Okada:2025lpl}, shifted hybrid inflation \cite{Jeannerot:2000sv, Kyae:2005fi, Khalil:2010cp, Lazarides:2020zof, Ahmed:2022thr, Afzal:2023cyp}, new inflation \cite{Senoguz:2004ky, Antusch:2008gw, Armillis:2012bs, Antusch:2013eca, Antusch:2014qqa, Rehman:2018gnr, Khan:2023snv}, $\mu$-hybrid inflation \cite{Dvali:1997uq, King:1997ia, Okada:2015vka, Wu:2016fzp, Rehman:2017gkm, Okada:2017rbf, Ahmed:2021dvo, Afzal:2022vjx, Ahmed:2024iyd, Ahmad:2025dds}, and D-term hybrid inflation \cite{Binetruy:1996xj, Halyo:1996pp}. As reported, these studies were motivated by the need to circumvent potential problems, for example, those associated with undesirable topological defects such as magnetic monopoles at the end of inflation or the $\mu$-problem of the minimal supersymmetric standard model (MSSM), etc. A usual approach followed in such models is to incorporate the breaking of a gauge group $G$ into the inflationary mechanism. $G$ is associated, for instance, with a grand unification theory (GUT), such as $SU(5)$, flipped $SU(5)$, Pati-Salam and $SO(10)$, or $U(1)_{B-L}$ extension of MSSM.

Due to a strong connection between SUSY and $U(1)_R$ or simply $R$-symmetry, manifested for example in the form of Nelson-Seiberg theorem \cite{Nelson:1993nf} that relates $R$-symmetry to SUSY-breaking mechanism, the interplay between the two have been a matter of great interest in studies involving SHI or its spin-offs. From the idea of imposing it to eliminate the undesirable self-couplings of the inflaton \cite{Dvali:1994ms} and to satisfy the slow-roll conditions \cite{Antoniadis:2017gjr}, violation of $R$-symmetry beyond the renormalizable level has been exploited to analyze its effects on inflationary predictions \cite{Civiletti:2013cra, Khalil:2018iip, Moursy:2020sit, Wan:2024wwu}. Another advantage of $R$-symmetry breaking at nonrenormalizable level is suppression of the so-called $\eta$-problem \cite{Copeland:1994vg, Stewart:1994ts, Dine:1995uk} in comparison with the normalizable level. Though the $R$-symmetry could be spontaneously broken due to quantum gravity effects or in the hidden sector, these studies utilize explicit $R$-symmetry breaking terms beyond the renormalizable in the superpotential for that purpose while keeping the symmetry exact at the renormalizable level. 

A recent study employed explicit $R$-symmetry breaking in the hidden sector, with its effects appearing at the renormalizable level, to realize phenomenologically viable $\mu$-hybrid inflation using a minimal Kähler potential \cite{Ahmad:2025dds}.
In addition, the same mechanism has been shown to support successful leptogenesis, a metastable cosmic string scenario and generation of gravitino dark matter abundance consistent with the observational constraints. In a separate work, SHI has been realized by introducing $R$-symmetry breaking interactions at the renormalizable level in the superpotential \cite{Wan:2025pcf}. Like earlier studies \cite{Civiletti:2013cra, Khalil:2018iip, Moursy:2020sit, Moursy:2021kst, Wan:2024wwu}, these models examine the implications of $R$-symmetry breaking through explicit terms in the superpotential, whether renormalizable or nonrenormalizable. Comparatively little attention has been devoted so far to the possibility of incorporating such effects through K\"{a}hler potential, a gap which we intend to address in this study. 
However, see Refs.~\cite{Pallis:2018xmt, Pallis:2020dxz, Lazarides:2023bjd, Pallis:2025epn} for related work, where mild $R$-symmetry breaking is assumed within the framework of hyperbolic K\"ahler geometry, successfully yielding an acceptably large mass for the potentially problematic $R$-axion.

An important motivation for considering $R$-symmetry breaking at a nonrenormalizable scale is its potential to resolve the problem of massless states that often arise in $R$-symmetric SUSY hybrid inflation models \cite{Barr:2005xya, Fallbacher:2011xg}. This problem typically appears when the inflaton acquires a vanishing vacuum expectation value at the time of symmetry breaking, leading to undesired massless modes. The presence of these fields can spoil gauge coupling unification. It has been demonstrated, for instance, that explicit $R$-symmetry violation beyond the renormalizable level can generate masses for right-handed neutrinos (RHNs) and down-type quarks within the context of flipped $SU(5)$ gauge group \cite{Civiletti:2013cra}. A similar strategy has recently been implemented in a realistic $SU(5)$ inflationary model, where $R$-symmetry breaking at the nonrenormalizable level enables the restoration of gauge coupling unification through the inclusion of a color octet and an electroweak triplet at an intermediate scale \cite{Ijaz:2023cvc}. For alternative solutions to this issue, see ref. \cite{Masoud:2019gxx} for a fine-tuned approach and ref. \cite{Antusch:2023mxx} for a solution without fine-tuning in the context of $SU(5)$ inflation.

In this study, we explore the role of explicit $R$-symmetry breaking (denoted as $\slashed{R}$) in an SHI scenario with the goal to make inflationary predictions consistent with the observational data, particularly the refinements reported in the Data Release 6 of Atacama Cosmology Telescope (ACT) \cite{ACT:2025tim}. 
Our analysis focuses on $\slashed{R}$ interactions that enter the inflationary dynamics through nonrenormalizable terms in the K\"{a}hler potential, rather than the more commonly examined contributions from the superpotential. This framework allows us to systematically quantify the influence of $\slashed{R}$ interactions on the scalar potential and, in turn, on the inflationary predictions of the model within the bounds set by recent cosmological observations. As an illustrative example, we present a symmetry-based argument that naturally accounts for the dominance of nonrenormalizable $\slashed{R}$ terms in the K\"{a}hler potential over those in the superpotential.

The contents of this study are arranged as follows. In section~\ref{sec:model}, we specify the forms of the superpotential and K\"{a}hler potential proposed to drive the inflation in addition to the discussion on the inflationary trajectory. Section~\ref{sec:inflation} offers a detailed description of non-minimal SHI within the proposed framework of the model. This includes a semi-qualitative analysis and comparison of the contributions to the scalar potential originated from the $\slashed{R}$ interactions of the superpotential and K\"{a}hler potential, apart from other factors. Following the specification of inflationary observables under the slow-roll framework in section~\ref{sec:slowroll}, and the discussion of reheating and non-thermal leptogenesis in section~\ref{sec:reheat}, a detailed analysis of the inflationary predictions, along with the assumptions and approximations underlying the quantitative results, is presented in section~\ref{sec:analysis}. The possibility of realizing the large-$r$ solution is explored in section~\ref{sec:larger}. In section~\ref{sec:conclusion}, we summarize the key points of the study and conclude.

\section{The Model}\label{sec:model}
We express the superpotential as a sum of renormalizable and nonrenormalizable components:
\begin{equation} {\label{eq:w}}
W = W_{\text{r}} + W_{\text{nr}} \,,
\end{equation}
where
\begin{equation} {\label{eq:wr}}
W_{\text{r}} = \kappa S (\Phi \bar{\Phi} - M^2) \,,
\end{equation}
is the most general renormalizable superpotential consistent with a gauge group $G$ and $R$-symmetry. Here, $S$ is a gauge singlet superfield with $R$-charge $R[S] = 1$, and $\Phi$, $\bar{\Phi}$ are a conjugate pair of superfields transforming under non-trivial representations of $G$, with a combined $R$-charge $R[\Phi \bar{\Phi}] = 0$. The nonrenormalizable part of the superpotential includes the leading-order Planck-suppressed $R$-violating terms, written as:
\begin{equation} {\label{eq:wnr}}
W_{\text{nr}} \supset \beta \frac{S ^4}{m_{P}} + \beta_1 \frac{S ^2 (\Phi \bar{\Phi})}{m_{P}} + \beta_2 \frac{(\Phi \bar{\Phi})^2}{m_{P}} \,.
\end{equation}
In these expressions, $\kappa$ is a dimensionless coupling that can be chosen positive without loss of generality, as its phase can be absorbed by a field redefinition of $S$. The coefficients $\beta$, $\beta_1$, and $\beta_2$ are also dimensionless. The parameter $M$ denotes the energy scale at which the gauge symmetry $G$ is spontaneously broken, and $m_P \simeq 2.4 \times 10^{18} \, \text{GeV}$ is the reduced Planck mass.

In SHI models, certain scalar components of the waterfall fields $\Phi$ and $\bar{\Phi}$ can exhibit flat directions. This issue, often referred to as the light waterfall field problem, arises due to the presence of an $R$-symmetry, which enforces the superpotential to be linear in the inflaton field $S$. As a consequence, only the symmetric combination $\Phi + \bar{\Phi}$ acquires a mass of order $\kappa M^2$, while the antisymmetric direction $\Phi - \bar{\Phi}$ remains either massless or receives a relatively small soft supersymmetry breaking mass of order $m_{3/2} \sim 10$~TeV. The presence of such light scalar states can interfere with gauge coupling unification. One effective way to lift these would-be massless modes is through the inclusion of explicit $R$-symmetry violating nonrenormalizable terms in the superpotential, such as $\beta_2 (\Phi \bar{\Phi})^2 / m_P$. While these terms do not affect the inflationary dynamics, since inflation occurs along the $\Phi = \bar{\Phi} = 0$ trajectory, they become relevant after inflation, particularly during the waterfall transition and in the vacuum where the fields develop nonzero expectation values. When $\Phi$ and $\bar{\Phi}$ acquire vacuum expectation values of order $M$, such operators generate mass corrections for scalar fields, typically of order $\beta_2 M^2 / m_P$, which lies in the range $10^{10}-10^{14}$~GeV for $M \sim 10^{16}$~GeV and $\beta_2 \sim 10^{-4}-1$. These intermediate scale masses help resolve tensions with gauge coupling unification. A concrete realization of this idea was recently demonstrated in a realistic $SU(5)$ inflationary model, where $R$-symmetry breaking at the nonrenormalizable level enabled the restoration of gauge unification by introducing a color octet and an electroweak triplet at intermediate mass scales \cite{Ijaz:2023cvc}.

We consider a general form of the K\"{a}hler potential composed of three parts: the canonical term $K_{\text{c}}$, an $R$-symmetry preserving nonrenormalizable term $K_{\text{nm}}$, and an $R$-symmetry violating nonrenormalizable part $K_{\text{nr}}$, such that:
\begin{equation} \label{eq:K}
K = K_{\text{c}} + K_{\text{nm}} + K_{\text{nr}},
\end{equation}
where
\begin{equation}
K_{\text{c}} = |S|^2 + |\Phi|^2 + |\bar{\Phi}|^2 ,
\end{equation}
is the canonical K\"{a}hler potential, and
\begin{equation} \label{eq:knm}
K_{\text{nm}} = \kappa_S \frac{|S|^4}{4 m_P^2} + \kappa_{SS} \frac{|S|^6}{6 m_P^4} ,
\end{equation}
contains leading-order $R$-symmetry preserving nonrenormalizable corrections. The $\slashed{R}$ contributions to the K\"{a}hler potential are given by the leading terms:
\begin{eqnarray} \label{eq:knr}
K_{\text{nr}} &\supset&
\frac{\alpha_1 |S|^2}{m_P} (S + S^\ast) + \frac{\alpha_2}{m_P} (S^3 + S^{\ast 3}) + \frac{\alpha_3 |\Phi|^2}{m_P} (S + S^\ast) + \frac{\alpha_4 |\bar{\Phi}|^2}{m_P} (S + S^\ast) \,.
\end{eqnarray}
Here, $\alpha_i$ (for $i = 1,2,3,4$), $\kappa_S$, and $\kappa_{SS}$ are dimensionless non-minimal couplings. Since the $R$-symmetry preserving terms in $K_{\text{nm}}$ appear at higher order in the Planck-suppressed expansion, their contributions to the inflationary dynamics are expected to be subdominant compared to those from the $R$-violating terms.

As mentioned in section~\ref{sec:introduction}, our objective is to investigate the effects of leading order nonrenormalizable $\slashed{R}$ interactions that appear through the K\"{a}hler potential on the inflationary predictions of the model. As such terms are also a part of superpotential~\eqref{eq:w}, we can simply assume without any knowledge of underlying physics that the $\slashed{R}$ contribution from $W$ is either missing or negligible. This translate to a desirable situation where we are only left with such part from K\"{a}hler potential to determine model predictions against the observable range of inflationary parameters. 

As discussed in the following, the inflationary trajectory in this model is defined by $\Phi = \bar{\Phi} = 0$. Consequently, all terms at the nonrenormalizable level in the superpotential and K\"{a}hler potential involving these superfields become irrelevant during inflation. Taking this into account along with the considerations above, we extract the relevant inflationary components of the superpotential from eq.~\eqref{eq:w} as:
\begin{equation} {\label{eq:win}}
W_{\text{inf}}=  \kappa S (\Phi \bar{\Phi} - M^2) + \beta \frac{S ^4}{m_{P}}\,,
\end{equation}
with the K\"{a}hler potential from eq.~\eqref{eq:K},
\begin{eqnarray} {\label{eq:kin}}
K_{\text{inf}} &=&  |S|^{2} +  |\Phi|^{2}+  |\bar{\Phi}|^{2} + \kappa_{S} \frac{\vert S\vert ^4}{4 m_{P}^{2}}+ \kappa_{SS} \frac{\vert S\vert ^6}{6 m_{P}^{4}} +  \frac{\alpha_{1} \vert S\vert ^2}{m_{P}}  \left(S + S^\ast\right) +  \nonumber \\
&&\frac{\alpha_{2}}{m_{P}}  \left(S^{3} +  {S^\ast}^{3}\right) \, .
\end{eqnarray}
The SHI model employing this form of $R$-symmetry breaking in the superpotential have been previously explored in detail, for instance in ref.~\cite{Civiletti:2013cra}, using a minimal canonical K\"{a}hler potential $K_c$.

We first calculate the global SUSY F-term scalar potential in 
the D-flat direction ($\Phi = \bar{\Phi}^{*}$) with $\beta$ term in $W_{\text{inf}}$ as
\begin{eqnarray} \label{eq:vfglobal}
V^{\text{global}}_{\text{F}} &\equiv& \sum_i \bigg|\frac{\partial W_{\text{inf}}}{\partial z_i}\bigg|^2  = \kappa^2 \bigg[\left|M^2 -|\phi|^2\right|^2 + 2|s|^2 |\phi|^2\bigg] + \frac{8 \beta}{m_P^2}\bigg[2 \beta  |s|^6 - \kappa m_P  |s|^3 \times\nonumber \\
    &&  (M^2 -|\phi|^2) \cos{(3\theta_S)}\bigg], 
\end{eqnarray}
where $z_i \in \{s, \phi, \bar{\phi}\}$, and $s$, $\phi$, and $\bar{\phi}$ denote the scalar components of the corresponding superfields $S$, $\Phi$, and $\bar{\Phi}$, respectively. The global SUSY vacuum corresponds to:
\begin{equation} \label{eq:gsm}
\langle s \rangle = 0, \quad \langle \phi \bar{\phi} \rangle = M^2 .
\end{equation}
During inflation, the real field $s$ and its angular component $\theta_S$ form a two-field inflaton system, while the conjugate pair $\Phi$, $\bar{\Phi}$ provides a constant vacuum energy of approximately $\kappa^2 M^4$, determined by the gauge symmetry breaking scale $M$. The full inflationary potential, expressed in terms of the fields $s$, $\theta_S$, and $\phi$, can be effectively reduced to a single-field model. This simplification is valid under the assumption that $s \gg M$ initially, ensuring sufficient inflation. In this regime, the fields $\theta_S$ and $\phi$ acquire heavy masses, due to dependence of the same on $s$, and rapidly settle into their respective minima at zero. As a result, $s$ remains the only dynamical field driving inflation.

The inflationary trajectory follows the local minimum at $|\phi| = 0$, which remains stable for large values of $s$. The inflaton field $s$ rolls slowly along this path until it reaches a critical value $s_c$, where the trajectory becomes unstable. This critical point is defined by the condition:
\begin{equation}
\frac{\partial^2 V}{\partial |\phi|^2} \bigg|_{|\phi|=0} = 0 .
\end{equation}
At $s = s_c$, the instability triggers a transition to one of the global minima at $|\phi| = \pm M$, marking the spontaneous breaking of the gauge symmetry $G$. In the absence of nonrenormalizable terms, the critical value is simply $s_c = M$. However, the inclusion of the $R$-violating term $S^4$ in $W_{\text{inf}}$ modifies this to:
\begin{equation}
s_c \simeq  M - \frac{2 \beta}{\kappa} \, \frac{M^2}{m_P}  .
\end{equation}
In the following section, we focus on scenarios where this $\beta$-term is either absent or highly suppressed. Consequently, the critical value remains at $ M$.

\section{Non-Minimal Supergravity Hybrid Inflation} \label{sec:inflation}
In this section, we incorporate additional contributions to the scalar potential, including radiative corrections ($\Delta V_{\text{1-loop}}$), soft SUSY-breaking terms ($V_{\text{soft}}$), and supergravity (SUGRA) corrections ($V_{\text{F}}$) \cite{Linde:1997sj}. The radiative corrections arise from SUSY-breaking due to the non-zero vacuum energy, which induces mass splitting between the fermionic and bosonic components of the relevant superfields \cite{Dvali:1994ms}. On the other hand, SUSY-breaking in the hidden sector leads to soft SUSY-breaking contributions to the scalar potential \cite{Buchmuller:2000zm,Senoguz:2004vu,Rehman:2009nq}.

To account for SUGRA corrections, we use the standard expression for the F-term scalar potential:
\begin{equation} \label{eq:SUGRA}
V_{\text{F}} =  e^{K/m_{P}^{2}} \left(K_{ij}^{-1} D_{z_{i}} W D_{z^{\ast}_{j}} W^\ast - \frac{3}{m_{P}^{2}}\vert W\vert ^2\right), 
\end{equation}
where
\begin{eqnarray}
K_{ij}  \equiv   \frac{\partial^2 K}{\partial z_{i} \partial z^{\ast}_{j}}\,,\,
D_{z_{i}} &W& \equiv \frac{\partial W}{\partial z_{i}}+\frac{1}{m_{P}^{2}} \frac{\partial K}{\partial z_{i}} W \,,\, 
 D_{z^{\ast}_{i}} W^\ast = \left(D_{z_{i}} W\right)^\ast \,.
\end{eqnarray}

The one-loop radiative corrections are computed using the Coleman-Weinberg formula \cite{Coleman:1973jx}:
\begin{equation} \label{eq:rcorrections}
\Delta V_{1-\text{loop}}(x)	=  \frac{(\kappa M)^4}{8 \pi^2} \mathcal{N} F(x),
\end{equation} 
where $x \equiv s/M$, $\mathcal{N}$ is the dimensionality of the representation of the superfields $\Phi$ and $\bar{\Phi}$, and the loop function $F(x)$ is given by:
\begin{eqnarray}
F(x) &=&\frac{1}{4}\bigg((x^4 + 1)\ln{\frac{x^4 - 1}{x^4}} + 2 x^2 \ln{\frac{x^2 + 1}{x^2 - 1}} + 2 \ln{\frac{\kappa^2 M^2 x^2 }{Q^2}} -3\bigg) \,,
\end{eqnarray}
with $Q$ being the renormalization scale. These one-loop corrections are computed within the framework of global SUSY, neglecting the effects of SUGRA and nonrenormalizable terms in $W$ and $K$.

The soft SUSY-breaking terms contribute to the scalar potential as:
\begin{equation} \label{eq:ssusy}
V_{\text{\text{soft}}}(x)	= a  \kappa m_{3/2}  M^3 x + m_{3/2}^{2} M^2 x^2 + \frac{b \beta m_{3/2}  M^4 x^4}{m_P}   ,
\end{equation}
where
\begin{equation} 
a = 2 \  \vert 2 - A \vert \cos[\theta_S + \arg (2-A)].
\end{equation}
The final term in eq.~\eqref{eq:ssusy} arises from the nonrenormalizable $S^4$ term in the superpotential \cite{Civiletti:2013cra}.

Using eq.~\eqref{eq:SUGRA}, \eqref{eq:rcorrections}, and \eqref{eq:ssusy}, the total scalar potential can be written as
\begin{equation}
V \approx V_{\text{F}} + \Delta V_{\text{1-loop}} + V_{\text{soft}}\,.    
\end{equation}
The scalar potential effectively receives $\slashed{R}$ contributions only through the terms arising in $K_{\text{inf}}$. As a result, along the inflationary trajectory where $\Phi = \bar{\Phi} = 0$, the potential simplifies to:
\begin{eqnarray} \label{eq:potential}
V (x,\theta_S) &=& V_0 \bigg[1- \bigg\{4 \alpha_{1} \cos(\theta_S) - \frac{ am_{3/2}  \ m_{P}}{\kappa M^2}  \bigg\} \left(\frac{M x}{m_{P}}\right) - \bigg\{ \kappa_{S} - \bigg(4 \alpha_{1} \cos(\theta_S)\bigg)^2 - \notag \\
&& \bigg(\frac{ m_{3/2}  \ m_{P} }{\kappa M^2}\bigg)^2 \bigg\} \left(\frac{M x}{m_{P}}\right)^2 - 4 \bigg\{ 2 \bigg(2 \alpha_{1}^{3} - \alpha_{2} + \frac{\beta  \ m_{P}^2 }{\kappa M^2} \bigg) \cos(3\theta_S) +\alpha_{1} \times  \notag \\
&&   \bigg(12 \alpha_{1} ^{2}-2 \kappa_{S} +1 \bigg)  \cos(\theta_S) \bigg\} \left(\frac{M x}{m_{P}}\right)^3 + \bigg\{\frac {1} {4}\bigg(4 \kappa_ {S}^2 - 7 \kappa_ {S} - 6 \kappa_ {SS} + 2 \bigg) +  \notag \\
&&  \bigg(4 \alpha_{1} \cos(\theta_S)\bigg)^2 \bigg\{  \bigg(4 \alpha_{1} \cos(\theta_S)\bigg)^2 - 3 \kappa_ {S} \, + \ 1 \bigg\} - \frac {32\alpha_ {1} \cos (\theta_S)^2} {\kappa M^2} \times \notag \\
&&  \bigg ( \kappa \alpha_ 2 M^2 - \beta \,  m_P^2 \bigg)   \bigg( 2\cos (2\theta_S) - 1 \bigg) + \frac{b \beta m_{3/2}m_P^3}{\kappa^2 M^4} \bigg\} \left (\frac {M x} {m_ {P}} \right)^4 - \frac {2 \beta \, m_P^2} {\kappa M^2}  \times \notag \\
&& \bigg\{2 \bigg( 2  \bigg(4 \alpha_{1} \cos(\theta_S)\bigg)^2 - 2 \kappa_S + 3 \bigg) \cos (3\theta_S) \left(\frac{M x}{m_{P}}\right)^5 + \mathcal{O} \left(\frac{M x}{m_{P}}\right)^6 \bigg\} + \notag \\
&& \frac{\kappa ^2 \mathcal{N}}{8 \pi^2}  F(x) \bigg] ,
\end{eqnarray}
where $V_0 \equiv \kappa^2 M^4$. Following the same reasoning as in section~\ref{sec:model}, the phase $\theta_S$ can be stabilized at zero before observable inflation through suitable initial conditions \cite{Senoguz:2004vu, Buchmuller:2014epa}. This allows us to treat $a$ as a constant of order unity.

The influence of explicit $R$-symmetry breaking and non-minimal $R$-symmetric contributions is encoded in the potential through the couplings $\beta, \, \alpha_i$ and $\kappa_S, \,\kappa_{SS}$, respectively. A preliminary analysis shows that higher-order terms of $\mathcal{O}(Mx/m_P)^5$ and beyond are strongly suppressed and can be safely neglected, especially for typical values such as $\kappa \sim 10^{-2}$, $M \sim 10^{15}$ GeV, and $m_{3/2} \sim$ TeV. 

It’s important to note that $\beta$ appears first at cubic order in field expansion, whereas $\alpha_1$ contributes linearly. Assuming $\theta_S = 0$, a comparison of these terms shows:
\begin{equation}
\beta  \, \Leftrightarrow \, \alpha_1 \left( \frac{\kappa}{x^2} \right) .  
\end{equation}
Since $\kappa < 1$ and $x > 1$ during inflation, the $\beta$-term typically dominates if we take $\beta \sim \alpha_1$. Therefore, to isolate and study the impact of the $\alpha_1$ term (arising from the K\"ahler potential), the $\beta$-term should be suppressed.

It is widely believed that global symmetries, including $U(1)_R$, are expected to be violated by quantum gravity effects (see, e.g., ref. \cite{Banks:2010zn}). Consequently, both the superpotential and K\"ahler potential are expected to receive $R$-violating corrections, typically suppressed by inverse powers of the Planck scale. To quantify the strength of this $R$-symmetry breaking, we introduce a small parameter $\epsilon \equiv (M_X/m_P)^p$, where $M_X$ is the scale at which $R$-symmetry breaking occurs and $p$ is an appropriate positive integer. In addition, we assume that the suppression of a given nonrenormalizable operator $O$ with $R$-charge $R_{O}$ scales as $\epsilon^{|R_{O} - 1|}$ in the superpotential and as $\epsilon^{|R_{O}|}$ in the K\"ahler potential. This structure is motivated by the Froggatt–Nielsen mechanism \cite{Froggatt:1978nt}, which is often invoked to explain hierarchical couplings in effective field theories. For instance, with $M_X \sim m_P/10$ and $p = 4$, we obtain $\epsilon \sim 10^{-4}$. This leads to $\alpha_{1,2,3,4} \sim \beta_{1,2} \sim 10^{-4}$ and $\beta \sim 10^{-12}$. Such a hierarchy naturally explains why the $S^4$ term in the superpotential is strongly suppressed compared to the $R$-violating terms with couplings $\alpha_{1,2}$ in the K\"ahler potential. We emphasize that our aim here is not to construct an explicit ultraviolet (UV) model that justifies this suppression from first principles, but rather to provide a phenomenologically motivated framework consistent with the general expectations of SUGRA.

With $\theta_S$ stabilized at zero, and neglecting the $\beta$ term along with other subdominant terms suppressed by powers of the field, the scalar potential in eq.~\eqref{eq:potential} reduces to:
\begin{eqnarray} \label{eq:potential2}
V (x) &\simeq& V_0 \bigg[1- \bigg(4 \alpha_{1} - \frac{ a \ m_{3/2}  \ m_{P}}{\kappa M^2}  \bigg) \left(\frac{M x}{m_{P}}\right) +	\bigg\{  16 \alpha_{1} ^{2} -\kappa_S+ \bigg(\frac{ m_{3/2}  \ m_{P} }{\kappa M^2}\bigg)^2 \bigg\}  \left(\frac{M x}{m_{P}}\right)^2  \notag \\
&-& 4 \bigg\{\alpha_{1}  \bigg(1+ 12 \alpha_{1} ^{2} -2 \kappa_S\bigg) + 2  \bigg(2 \alpha_{1}^{3} - \alpha_{2} \bigg) \bigg\}   \left(\frac{M x}{m_{P}}\right)^3 + \bigg\{\frac {1} {4}\bigg(4 \kappa_ {S}^2 - 7 \kappa_ {S} - 6 \kappa_ {SS} + 2 \bigg)   \notag \\
&+&  16 \alpha_{1} ^{2} \bigg(  16 \alpha_{1} ^{2} -3 \kappa_S + 1 \bigg) - 32\alpha_ {1} \alpha_2 \bigg\} \left (\frac {M x} {m_ {P}} \right)^4  +  \frac{\kappa ^2 \mathcal{N}}{8 \pi^2}  F(x) \bigg] .
\end{eqnarray}
It is worth noting that the critical point for the scalar potential with SUGRA corrections remains unchanged and is found to be $x_c \simeq 1$ (or, equivalently, $s_c \simeq M$) with additional terms having a negligible effect on its value. 
\section{Slow-Roll Approximation}\label{sec:slowroll}
To extract the quantitative inflationary observables from the scalar potential, we use the slow-roll approximation defined by the so-called slow-roll parameters $\epsilon$, $\eta$, and $\zeta$. These parameters are given by
\begin{align}
\epsilon  = \frac{1}{2} m_P^2 &
\left(\frac{ V_{\sigma}}{V}\right)^2 \,\,, \,\,\,
\eta = m_P^2
\left( \frac{V_{\sigma \sigma}}{V} \right) \,\,, \,\,\, 
\zeta^2  = m_P^4
\left( \frac{V_{\sigma} V_{\sigma \sigma \sigma} }{V^2}\right).
\label{eq:slowroll}
\end{align}
Here $V_{\sigma} \equiv \partial V / \partial \sigma$ and similarly multiple derivatives are taken with respect to canonically-normalized field (the inflaton), which is related to $x$ through $\sigma \equiv\sqrt{2}  s = \sqrt{2}  M x$. These parameters relate in leading order to the standard inflationary observables including the scalar spectral index, $n_s$, the tensor-to-scalar ratio, $r$, and the running of the scalar spectral index, $\alpha_{s}\equiv dn_s / d \ln k$. The following expressions stipulate this relationship:
\begin{align} \label{eq:rns}
n_s&\simeq 1-6\,\epsilon+2\,\eta\,\,\,, \,\,\, r\simeq 16\,\epsilon \,\,\,, \,\,\,
\alpha_{s} \simeq 16\,\epsilon\,\eta
-24\,\epsilon^2 - 2\,\zeta^2.
\end{align}
Substituting $V(x)$ from eq.~\eqref{eq:potential2} in the slow-roll expressions, $n_s$,  $r$ and $\alpha_{s}$ can be approximated as
\begin{eqnarray} \label{eq:ns2}
n_s &\simeq& 1-2 \kappa_S -24  \alpha_{1} ^{2} - 24 (\alpha_{1} + \kappa_S \alpha_{1} -2 \alpha_{2}) \left(\frac{M x_0}{m_{P}}\right) - 3(7 \kappa_S +  6\kappa_{SS}- 2)
 \left(\frac{M x_0}{m_{P}}\right)^2 -\notag \\ 
&&  \frac{\kappa ^2 \mathcal{N} m_{P}^2}{8 M^2 \pi^2} |F''(x_0)| \, ,
\end{eqnarray}
\begin{eqnarray} \label{eq:ralphas2}
r &\simeq& 64 \alpha_{1} ^{2} + 64 \kappa_S \alpha_{1}  \left(\frac{M x_0}{m_{P}}\right)  + \frac{\kappa ^2 \mathcal{N} m_{P}  F'(x_0)}{M \pi^2}\bigg( 4 \alpha_{1} +   \frac{\kappa ^2 \mathcal{N} m_{P} }{16 M \pi^2} F'(x_0) \bigg),  \\ 
\alpha_{s} &\simeq& -48 \alpha_{1}^2 + 96 \alpha_{1} \alpha_{2} - \frac{\kappa ^2 \mathcal{N} m_{P}^2  F'''(x_0)}{128 M^4 \pi^4}\bigg\{\kappa ^2 \mathcal{N} m_{P}^2  F'(x_0) - 16 M \pi^2 m_{P} \bigg(2  \alpha_{1} +  \notag \\
&& \kappa_S \left(\frac{M x_0}{m_{P}}\right) \bigg)\bigg\} ,
\end{eqnarray}
where primes are the derivatives with respect to $x$ and $x_0 \equiv x(k_0)$ is the field value at the pivot scale $k_0 = 0.05$ $\text{Mpc}^{-1}$. While the expression for the scalar spectral index could be further simplified by omitting subdominant terms---for instance, those arising from higher-order couplings or field power---this form is retained here to illustrate the contributions of $\kappa_S$ and $\kappa_{SS}$. On the other hand, since the tensor-to-scalar ratio lacks any explicit linear dependence on the couplings, it is clear that $r \lesssim 10^{-5}$ for $\alpha_{1}\lesssim 10^{-3}$.

The amplitude of the scalar power spectrum, $A_s$, is given by,
\begin{align}
A_{s}(k_0) = \frac{1}{24\,\pi^2\,\epsilon}
\left( \frac{V}{m_P^4}\right)\bigg|_{x=x_0},
\end{align}
where $A_s(k_0) = 2.137 \times 10^{-9}$ as measured by Planck \cite{Planck:2018vyg, Planck:2018jri}. In a similar fashion to above, we can approximate the power spectrum in this case as
\begin{eqnarray} \label{Asexp}
A_{s} &\simeq& \frac{\kappa^2}{24\,\pi^2}
\left( \frac{ M}{m_P}\right)^4\Bigg[4 \alpha_{1} ^{2} + 4 \kappa_S \alpha_{1}  \left(\frac{M x_0}{m_{P}}\right)  +  \frac{\kappa ^2 \mathcal{N} m_{P}  F'(x_0)}{4 M \pi^2}\bigg(\alpha_{1} + \notag \\
&&\frac{\kappa ^2 \mathcal{N} m_{P} }{64 M \pi^2} F'(x_0) \bigg) \Bigg]^{-1}. 
\end{eqnarray}

The amount of observable inflation is given by the number of e-folds $N_0$ before the end of inflation, and is written in terms of $V$ as,
\begin{align}\label{eq:Ngen}
N_0 = 2 \left( \frac{M}{m_P}\right) ^{2}\int_{x_e}^{x_{0}}\left( \frac{V}{	V'}\right) dx,
\end{align}
where $x_e$ is the field value at the end of inflation, which for numerical analysis, is fixed either by the breakdown of the slow-roll approximation [$\eta(x_e)=-1$], or by a `waterfall' destabilization occurring at $x_e = 1$. 
\section{Reheating and Leptogenesis}\label{sec:reheat}
While a detailed analysis of reheating requires specifying the gauge symmetry group $G$, we outline here a general and well-motivated scenario in which the observed baryon asymmetry of the universe is explained through leptogenesis within SHI models \cite{Senoguz:2003hc}. Once the gauge symmetry is fixed, both reheating and leptogenesis can be naturally incorporated by including suitable terms in the superpotential of the form:
\begin{equation}
W \supset \gamma_{ij} f(\Phi, \bar{\Phi}) N_i N_j + y_{ij} N_i L_j H  ,
\end{equation}
where the function $f(\Phi, \bar{\Phi})$ depends on the specific model and may be taken as
\begin{equation}
f(\Phi,  \bar{\Phi}) =  \Phi \,\,\text{ or }\,\,  \bar{\Phi}\,\,  \text{ or }\,\, \frac{\Phi \bar{\Phi}}{m_P}.
\end{equation}
Here, $i, j$ are generation indices; $\gamma_{ij}$ and $y_{ij}$ denote dimensionless and Yukawa couplings, respectively; and $N_i$, $L_j$, and $H$ represent the RHN superfields, lepton doublets, and MSSM Higgs doublet superfield in that order. When the waterfall fields $\Phi$ or $\bar{\Phi}$ acquire a vacuum expectation value, the resulting symmetry breaking generates masses for the RHNs. Moreover, the scalar components of these fields can decay into RHNs, providing the lepton-number-violating interactions required for successful leptogenesis.
The RHN masses are typically of the form:
\begin{equation}
   M_i = \gamma_i M \,\, \text{ or }\,\, \gamma_i \frac{M^2}{m_P},
\end{equation}
depending on the structure of the model, where $\gamma_i$ are the eigenvalues of $\gamma_{ij}$. The mass of the heaviest RHN is expected to be around $10^{14}$ GeV.

The reheating temperature $T_R$ is related to the inflaton decay width $\Gamma_{\text{inf}}$ by
\begin{equation} \label{eq:tr}
T_R=\left(\dfrac{90}{\pi^{2}g_{*}}\right)^{1/4}\left(\Gamma_\text{inf} \, m_{P}\right)^{1/2},
\end{equation}
where $g_* = 228.75$ for the MSSM. The decay width $\Gamma_{\text{inf}}$ depends on the matter content and interactions determined by the gauge symmetry. For standard thermal history, the number of e-folds $N_0$ between the horizon exit of the pivot scale and the end of inflation is given by \cite{Kolb:1990vq, Liddle:2003as}:
\begin{align}\label{eq:efolds_in_tr}
N_0=53+\dfrac{1}{3}\ln\left[\dfrac{T_R}{10^9 \text{ GeV}}\right]+\dfrac{2}{3}\ln\left[\dfrac{\sqrt{\kappa}\,M}{10^{15}\text{ GeV}}\right].
\end{align}
The inflaton decay rate into RHNs is given by
\begin{equation}\label{eq:gammainf}
 \Gamma_\text{inf} = \frac{1}{8\pi} \frac{M_N^2}{M^2} m_{\text{inf}},   
\end{equation}
where $m_{\text{inf}} = \sqrt{2} \kappa M$ is the inflaton mass and $M_N = M_1$ denotes the mass of lightest neutrino which participates in leptogenesis. A successful generation of lepton asymmetry compatible with observations can be achieved for values of $\kappa$ and $M$ explored in our numerical analysis \cite{Senoguz:2003hc}. We assume a reheating temperature of $T_R \simeq 10^9$ GeV, which is consistent with the upper bound from gravitino overproduction constraints \cite{Khlopov:1984pf, Ellis:1984eq, Kawasaki:2008qe, Kawasaki:2017bqm}, namely $T_R \lesssim 2 \times 10^9$ GeV. This bound corresponds to a gravitino mass of $m_{3/2} \sim 10$ TeV, which we adopt throughout our numerical study. Such a mass is also compatible with a SUSY spectrum within the reach of current and future collider experiments, including the LHC and FCC \cite{FCC:2018vvp}.

We can quantitatively assess the consistency of our model to successfully predict leptogenesis by employing the observed baryon asymmetry $n_\mathrm{B} /n_\gamma$. 
$\Gamma_\text{inf}$ contributes to the observed baryon asymmetry through the sphaleron process \cite{Kuzmin:1985mm,Fukugita:1986hr,Khlebnikov:1988sr}, which can be estimated, with the assumption that the inflaton solely decays into RHN according to eq.~\eqref{eq:gammainf}, as
\begin{align}\label{eq:bphr}
\frac{n_{B}}{n_{\gamma}}\simeq -1.84 \,\varepsilon_L  \frac{T_R}{m_\text{inf}} \delta_{\text{eff}} \,,
\end{align} 
where $\delta_{\text{eff}}$ is the $CP$-violating phase factor, with the constraint $|\delta_{\text{eff}}|\leq 1$, and $\varepsilon_L$ is the lepton asymmetry factor. Assuming hierarchical neutrino masses $M_1 \ll M_2 <M_3$ \cite{Rehman:2018gnr}, $\varepsilon_L$ is given by 
\begin{eqnarray}
\varepsilon_L  \simeq -\frac{3}{8\pi}  \frac{\sqrt{\Delta m_{31}^{2}} M_N}{\langle H_{u}\rangle^{2}}\,,
\end{eqnarray} 
where the atmospheric neutrino mass squared difference is $\Delta m_{31}^{2}\approx 2.6 \times 10^{-3}$ eV$^{2} $  and $\langle H_{u}\rangle \simeq 174$ GeV in the large $\tan\beta$ limit.Utilizing the observed ratio $n_\mathrm{B} /n_\gamma = (6.12 \pm 0.04) \times 10^{-10}$ \cite{ParticleDataGroup:2020ssz}, the kinematic bound, $2 M_{2,3} > m_\text{inf} \geq 2M_N$, the maximum value of CP violating phase, $|\delta_{\text{eff}}| = 1$, as well as the suppression of the washout factor of lepton asymmetry (achievable through $M_N \gg T_R$), we determine the range for $M_N$ that indicate comparability of the parametric space of our model with consistent leptogenesis. The numerical results are discussed in section~\ref{sec:analysis} below. Data presented in table~\ref{tab:values} indicate sufficient suppression of washout effects with $4 \, T_R \lesssim M_N \lesssim  34 \, T_R$.
\section{Numerical Analysis}\label{sec:analysis}
Before presenting the numerical results, it is useful to identify the effective number of parameters that play a crucial role in making the inflationary predictions consistent with cosmic microwave background (CMB) observations. Based on the preliminary discussion in section~\ref{sec:inflation}, the model’s key independent parameters are the symmetry breaking scale $M$, and the dimensionless couplings $\kappa$, $\kappa_S$, $\kappa_{SS}$, $\alpha_1$, and $\alpha_2$. Additional parameters required for numerical evaluation include the renormalization scale $Q$, the gravitino mass $m_{3/2}$, a dimensionless constant $a$, and the dimensionality $\mathcal{N}$ of the waterfall fields.

The value of $\mathcal{N}$ is determined by the choice of the gauge group $G$, and it plays a central role in computing radiative corrections. Since a detailed treatment of specific gauge groups lies beyond the scope of this work, we primarily adopt $\mathcal{N} = 1$, which corresponds, for instance, to the $U(1)_{B-L}$ gauge symmetry. Nevertheless, we briefly explore alternative choices of $G$ to estimate their impact on radiative corrections. These include $\mathcal{N} = 2, 8, 10, 16, 24$, associated respectively with the following gauge groups: left-right symmetric $SU(3)_c \times SU(2)_L \times SU(2)_R \times U(1)_{B-L}$, Pati–Salam $SU(4)_c \times SU(2)_L \times SU(2)_R$, flipped $SU(5)$, $SO(10)$, and Georgi–Glashow $SU(5)$.

For the numerical analysis, we fix the renormalization scale at $Q = M$, and set the gravitino mass to $m_{3/2} = 10\, \text{TeV}$, making the low-energy supersymmetric spectrum accessible at colliders such as the LHC and FCC. As for the dimensionless couplings from the K\"ahler potential, we assume them to be of the same order, i.e., $\alpha_1 = \alpha_2 = \kappa_S = \kappa_{SS}$. With this assumption, we proceed to investigate their role in shaping the inflationary predictions.

First, we discuss the impact of the $R$-symmetric non-minimal K\"ahler terms governed by the couplings $\kappa_S$ and $\kappa_{SS}$. These couplings contribute to the scalar potential [in eq.~\eqref{eq:potential2}] at quadratic and quartic orders in the field expansion, corresponding to terms suppressed by $(s/m_P)^2$ and $(s/m_P)^4$, respectively. Assuming $\kappa_S$ and $\kappa_{SS}$ are of the same order as the $\alpha_i$ couplings, their effects remain Planck-suppressed and relatively minor in the regime of small field excursions ($s \ll m_P$).
In contrast, the $R$-symmetry violating term associated with $\alpha_1$ appears at linear order in field suppression and thus dominates the scalar potential in this regime. These considerations justify neglecting the non-minimal Kähler corrections when all couplings are small and the inflaton field remains well below the Planck scale.
However, as the field value approaches $m_P$, the influence of $\kappa_S$ and $\kappa_{SS}$ becomes significant and must be taken into account. For a detailed comparison of the roles played by these couplings in an $R$-symmetric, non-minimal SHI model, see ref.~\cite{Shafi:2010jr}.

Next, we assess the significance of the soft SUSY-breaking term characterized by the gravitino mass $m_{3/2}$ and constant $a$. This term originates from the spontaneous breaking of $R$-symmetry in the hidden sector and enters the scalar potential as a soft SUSY-breaking contribution. Notably, both $R$-symmetry violating parameters, $\alpha_1$ and $a \,m_{3/2}$, contribute at linear order in the field expansion. By comparing their contributions with $a = 1$, we find that the term arising from the K\"ahler potential dominates when $\alpha_1 > m_{3/2}\, m_P / (4\kappa M^2)$. For symmetry breaking scales in the range $M \simeq (1-5) \times 10^{15}$~GeV, this condition translates to $\alpha_1 \kappa > 2 \times 10^{-8}$ to $10^{-9}$, assuming $m_{3/2} = 10\, \text{TeV}$.

\begin{figure} [!]
\centering
\subfloat[\label{fig:fixed_N_varying_alpha1_nsvk}]
{{\includegraphics[width=0.47\textwidth]{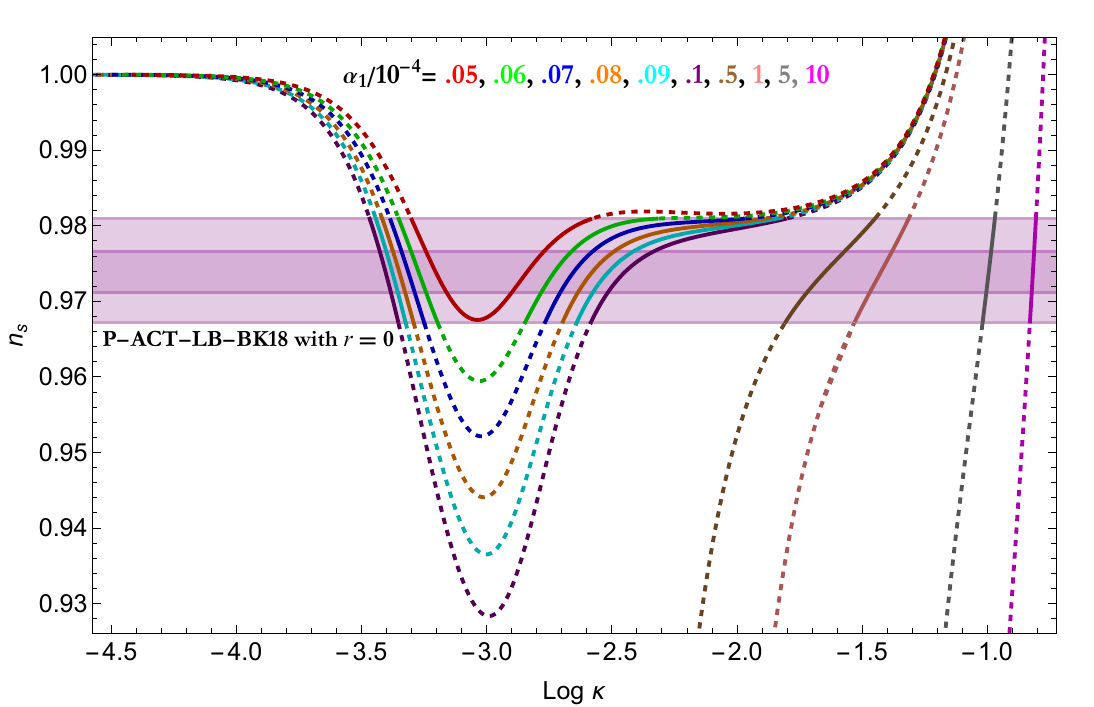}}}
\quad
\subfloat[\label{fig:fixed_N_varying_alpha1_rvns}]
{{\includegraphics[width=0.47\textwidth]{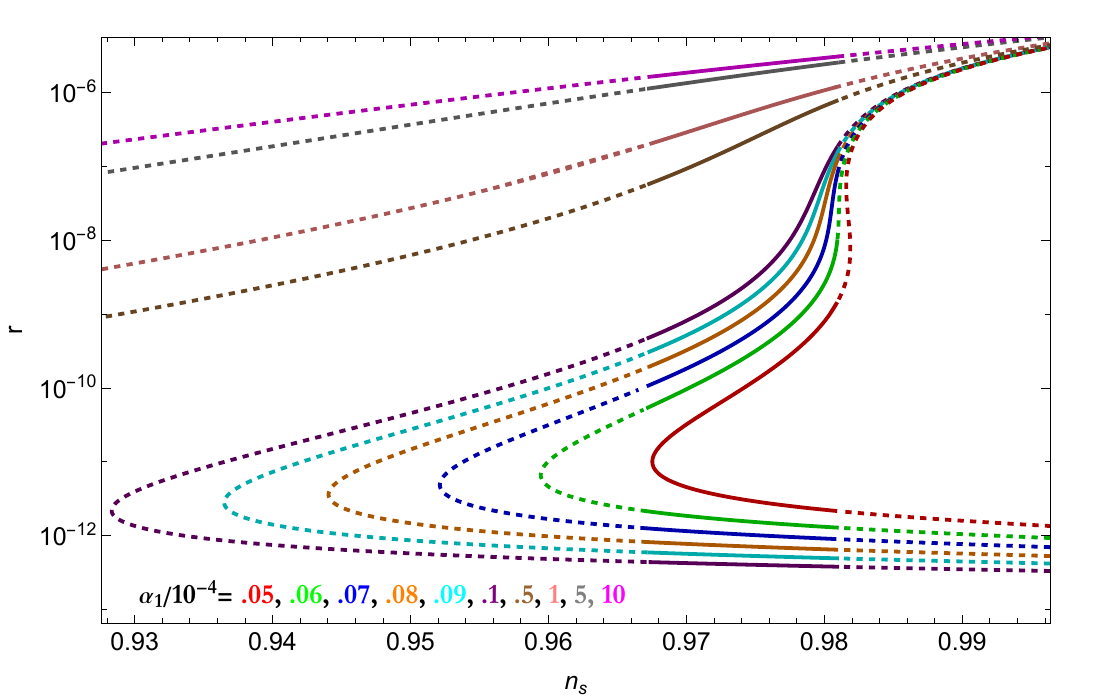}}} \\
\subfloat[\label{fig:fixed_N_varying_alpha1_Mvk}]
{{\includegraphics[width=0.47\textwidth]{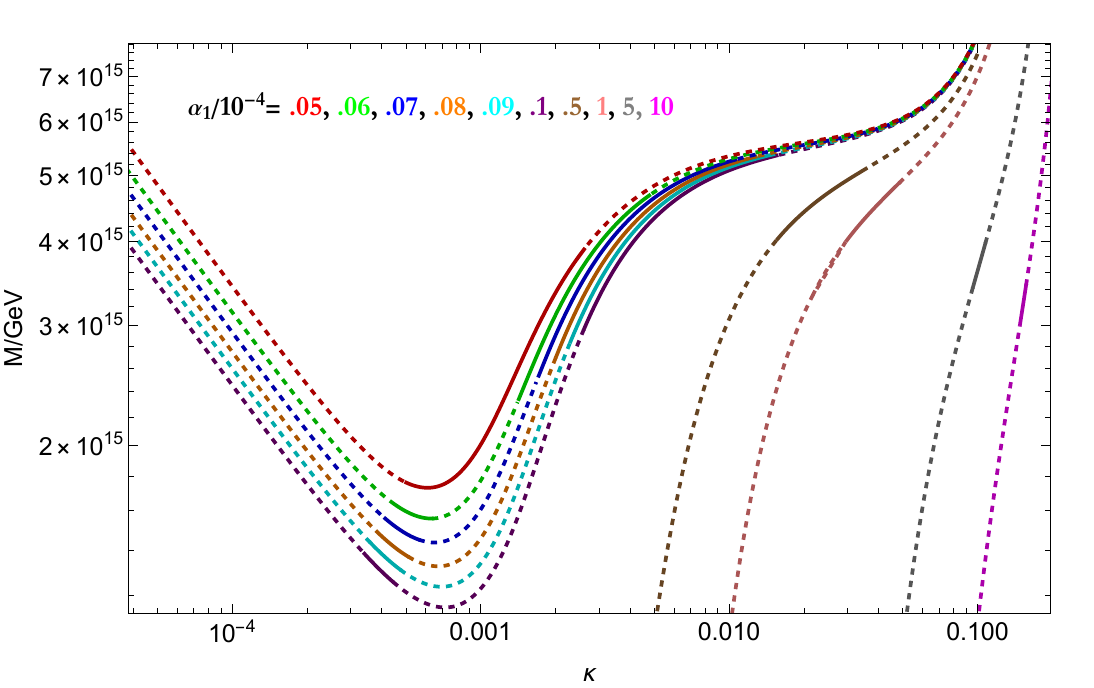}}}
\caption{Results of SHI as outlined in section~\ref{sec:analysis}, assuming $\kappa_S = \kappa_{SS} = \alpha_2 \equiv \alpha_1$, $\mathcal{N} = 1 = a$, reheating temperature $T_R \simeq 10^9$ GeV, and gravitino mass $m_{3/2} = 10$ TeV. From top to bottom, the plots show the scalar spectral index $n_s$, the tensor-to-scalar ratio $r$, and the symmetry breaking scale $M$ as functions of $\kappa$, with fixed values of $\alpha_1$. Each curve corresponds to a different $\alpha_1$ value, as labeled. The purple bands in figure~\ref{fig:fixed_N_varying_alpha1_nsvk} indicate the $1\sigma$ and $2\sigma$ confidence intervals for $n_s$ based on ACT Data Release 6 (P-ACT-LB-BK18 with $r=0$). Solid segments of the curves identify regions consistent with the $2\sigma$ bounds on $n_s$, while dashed segments show parameter values allowed by the model but outside the preferred observational range.}
\label{fig:n1a1_fixalpha1}
\end{figure}

\begin{figure}[!]
\centering
\subfloat[\label{fig:fixed_N_varying_k_nsvalpha1}]
{{\includegraphics[width=0.47\textwidth]{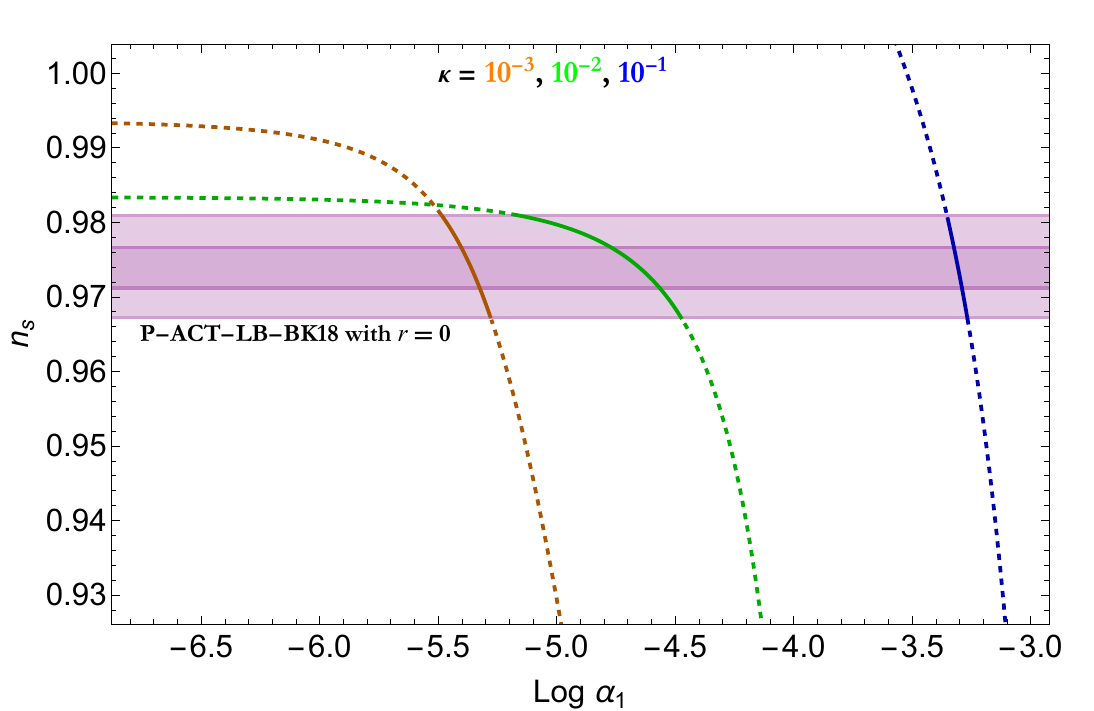}}}
\quad
\subfloat[\label{fig:fixed_N_varying_k_Mvalpha1}]
{{\includegraphics[width=0.47\textwidth]{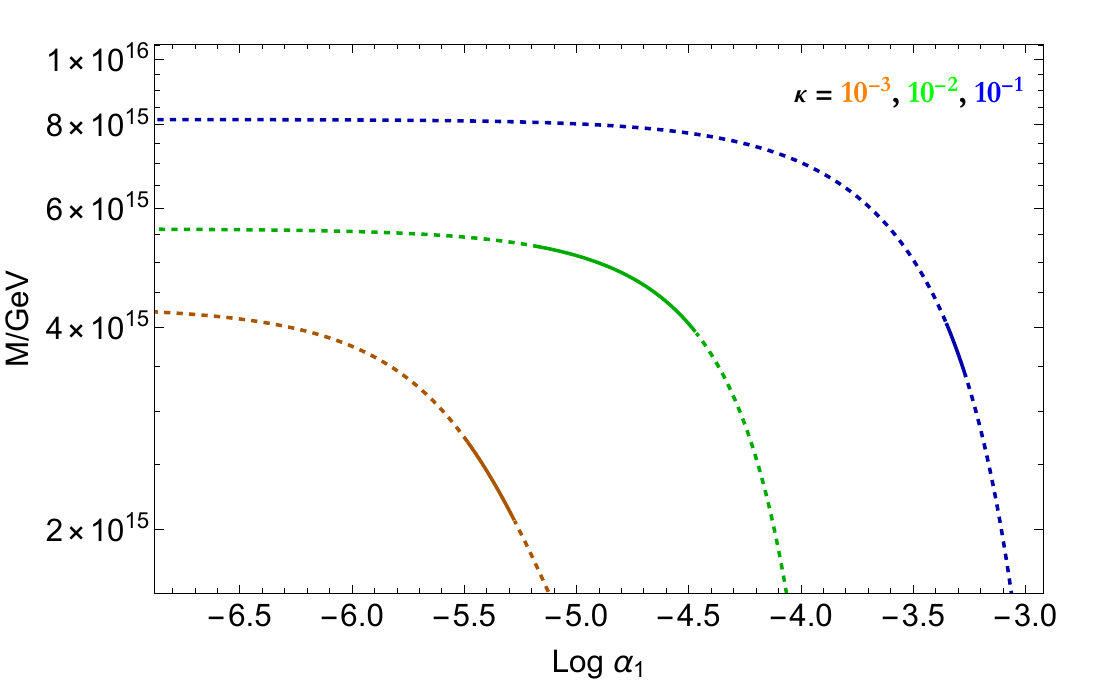}}}
\caption{Using the same inflationary framework and formatting as in figure~\ref{fig:n1a1_fixalpha1}, these plots illustrate how $n_s$ and $M$ vary with $\alpha_1$ for selected fixed values of $\kappa$. The constraints from P-ACT-LB-BK18 are applied, and figure~\ref{fig:fixed_N_varying_k_nsvalpha1} highlights the viable range of $\alpha_1$ consistent with observational bounds.}
\label{fig:n1a1_fixk}
\end{figure}

\begin{figure}[t!]
\centering
\subfloat[\label{fig:fixed_alpha1_varying_N_nsvk}]
{{\includegraphics[width=0.47\textwidth]{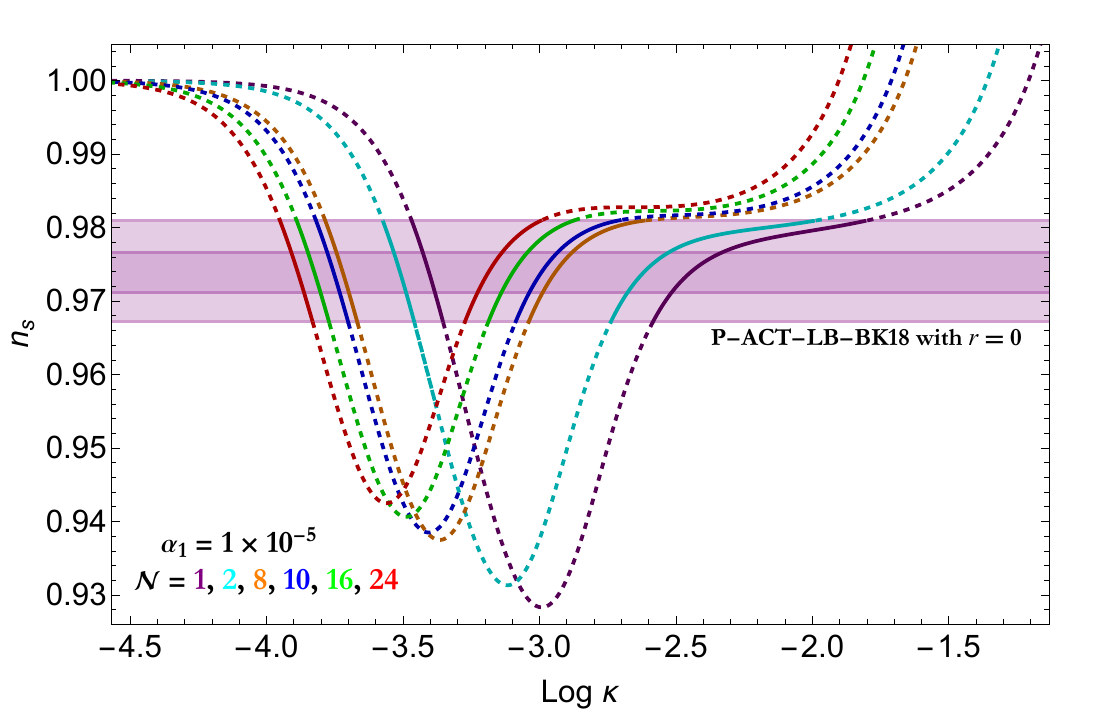}}}
\quad
\subfloat[\label{fig:fixed_alpha1_varying_N_rvk}]
{{\includegraphics[width=0.47\textwidth]{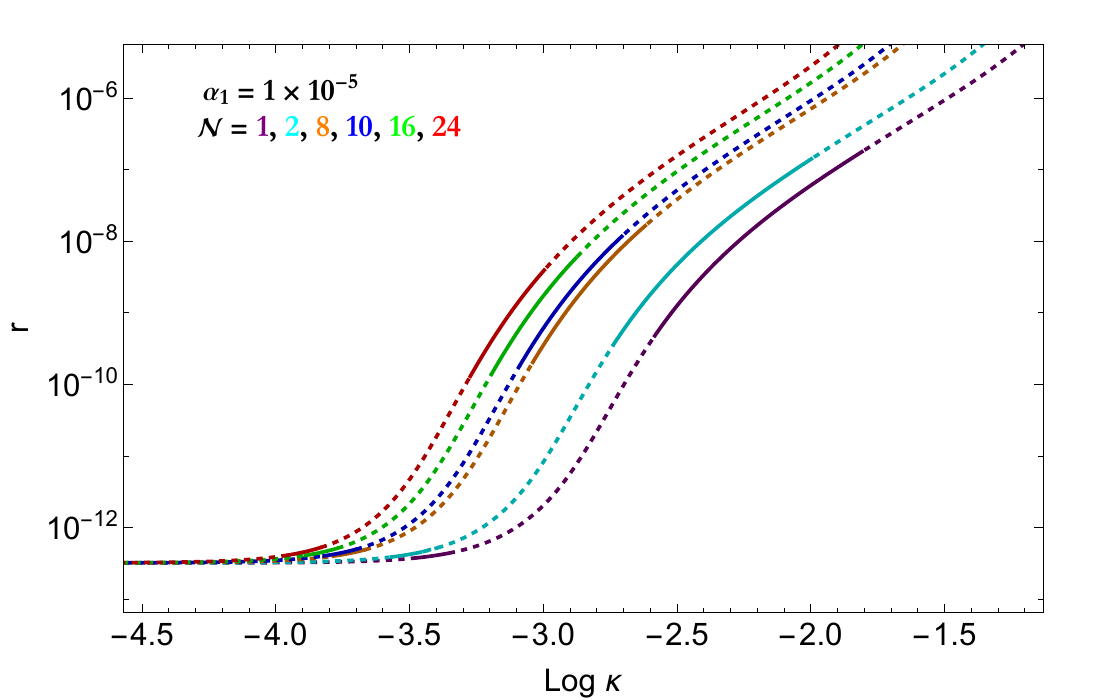}}}
\caption{Effects of increasing $\mathcal{N}$ on the inflationary predictions with $\alpha_1$ fixed at $10^{-5}$, while all the assumptions and legend style stay as before. The values selected for $\mathcal{N}$ correlate with some of the well-known gauge groups, as mentioned in section~\ref{sec:analysis}.}
\label{fig:varing_N_fixalpha1}
\end{figure}

Quantitative predictions for key parameters such as $\alpha_1$, $\kappa$, and $M$, along with the resulting CMB observables $n_s$ and $r$, are illustrated in figure~\ref{fig:n1a1_fixalpha1}, figure~\ref{fig:n1a1_fixk}, and figure~\ref{fig:varing_N_fixalpha1}. These results are based on the inflationary potential in eq.~\eqref{eq:potential2}, and are compared against observational data from the recent Data Release 6 of ACT \cite{ACT:2025tim}, particularly using the P-ACT-LB-BK18 bounds on $n_s$ as a reference.
To determine the number of e-folds $N_0$, we employ eq.~\eqref{eq:efolds_in_tr} using a fixed reheating temperature $T_R$, and we adopt the observationally measured amplitude of the scalar power spectrum as discussed in section~\ref{sec:inflation}. Inflation is assumed to end via the waterfall transition, corresponding to the condition $x_e = 1$. Under these constraints, only two parameters remain free to vary. Our predictions are obtained by fixing $\alpha_1$ and scanning over values of $M$ or $\kappa$, as shown in the panels of figure~\ref{fig:n1a1_fixalpha1}.

To assess the inflationary predictions for a parametric space where the effects of $\kappa_S$, $\kappa_{SS}$ and $\alpha_2$ are suppressed, while assuming them equal to $\alpha_{1}$, we simplify the potential in eq.~\eqref{eq:potential2}: 
\begin{eqnarray} \label{eq:case1potential}
V (x) & \simeq & V_0 \bigg[1- \bigg(4 \alpha_{1} - \frac{  m_{3/2}  m_{P}}{\kappa M^2}  \bigg) \left(\frac{M x}{m_{P}}\right)  +	  \frac {1} {2} \left (\frac {M x} {m_ {P}} \right)^4 +  \frac{\kappa ^2 \mathcal{N}}{8 \pi^2}  F(x) \bigg]. 
\end{eqnarray}
This form of potential leaves us with only three free parameters, $\kappa$, $\alpha_{1}$ and $M$ or $V_0 = \kappa^2 M^4$. For small field values with $\kappa < 10^{-2}$ where quartic SUGRA corrections can be neglected, inflationary predictions are controlled by the linear term and radiative corrections. Linear term with $\alpha_1 > m_{3/2}\, m_P / (4\kappa M^2)$ partially cancel the radiaitve correction in order to generate the measured value of the amplitude of the scalar power spectrum, $A_s \sim 10^{-9}$ [see eq.~\eqref{Asexp}] whereas the red-tilted scalar spectral index is achieved from the radiaitve corrections. This similar observation is discussed in \cite{Rehman:2009nq} where $a=-1$ is considered with minimal canonical K\"ahler potential without $\alpha_1$ term. As emphasized recently, this linear term is very important in making the predictions of SHI model consistent with the recent ACT Data Release 6, Planck 2018 and LB-BK18 \cite{Rehman:2025fja}.  

As depicted in figure~\ref{fig:n1a1_fixalpha1}, for a typical range of $M$, two distinct solution branches appear consistent with the observed value of $n_s$, particularly near $\kappa \sim 10^{-3}$. At smaller $\kappa$, where the soft SUSY-breaking term dominates, the scalar spectral index tends toward unity, consistent with findings in \cite{Rehman:2009nq}. The branch at larger $\kappa$ values is driven primarily by the $\alpha_1$ coupling, which originates from the $R$-symmetry-violating term in $K_{\text{nr}}$. Interestingly, these degenerate solution branches vanish when the $\slashed{R}$ terms are removed and $a = 1$ is assumed.
For the case under consideration, we find that the Planck-preferred central value $n_s \simeq 0.965$ can be achieved within the parameter ranges $9 \times 10^{-4} \lesssim \kappa \lesssim 0.35$ and $5.4 \times 10^{-6} \lesssim \alpha_1 \lesssim 3.8 \times 10^{-3}$. This window expands slightly if we instead match the ACT central value $n_s \simeq 0.974$, yielding $9.6 \times 10^{-4} \lesssim \kappa \lesssim 0.64$ and $4.21 \times 10^{-6} \lesssim \alpha_1 \lesssim 0.01$. In all cases, the running of the spectral index remains negligible, with $|\alpha_s| \lesssim 10^{-3}$. The ACT-compatible values for fixed $\alpha_1$ scenarios are summarized in table~\ref{tab:values}.

 \begin{table}[t]
 \centering
\resizebox{\linewidth}{!}{

\begin{tabular}{|ccccccccc|}
\hline
$\alpha_1$&$\kappa$&$r$&$|\alpha_{s}|$&$M / 10^{15}\text{ GeV}$&$m_{\textbf{inf}} / 10^{14}\text{ GeV}$&$M_N / 10^{10}\text{ GeV}$&$s_0/m_P$&$N_0$\\
\hline
\hline
$1.0 \times 10^{-3}$&$(1.5-1.6)\times 10^{-1}$&$\lesssim 3.0\times 10^{-6}$&$\lesssim  1.0 \times 10^{-3}$&$3.0 -3.4$&$6.3-7.7$&$3.0-3.4$&$(6.8-7.6)\times 10^{-2}$&$53.6-54.6$\\
$5.0 \times 10^{-4}$&$(1.0-1.1)\times 10^{-1}$&$\lesssim 2.5\times 10^{-6}$&$\lesssim  8.6 \times 10^{-4}$&$3.4-4.0$&$4.7-6.1$&$3.0-3.4$&$(4.9-5.9)\times 10^{-2}$&$53.0-53.2$\\
$1.0 \times 10^{-4}$&$(3.0-5.0)\times 10^{-2}$&$\lesssim 1.2\times 10^{-6}$&$\lesssim  5.5 \times 10^{-4}$&$4.0-4.9$&$1.7-3.4$&$2.5-3.3$&$(1.8-3.3)\times 10^{-2}$&$52.8-53.0$\\
$5.0 \times 10^{-5}$&$(2.0-4.0)\times 10^{-2}$&$\lesssim 7.8\times 10^{-7}$&$\lesssim  4.7 \times 10^{-4}$&$4.0-5.1$&$0.9-2.6$&$2.1-3.1$&$(0.9-2.6)\times 10^{-2}$&$52.5-53.0$\\
$1.0 \times 10^{-5}$&$(0.3-2.0)\times 10^{-2}$&$\lesssim 1.8\times 10^{-7}$&$\lesssim  3.7 \times 10^{-4}$&$3.0-5.4$&$0.1-1.2$&$1.1-2.6$&$(0.2-1.2)\times 10^{-2}$&$51.7-52.7$\\
&$(3.4-4.4)\times 10^{-4}$&$\lesssim 4.4\times 10^{-13}$&$\lesssim  1.7 \times 10^{-4}$&$1.3-1.4$&$.007-.008$&$0.4$&$(5.3-5.7)\times 10^{-4}$&$50.6$\\
$5.0 \times 10^{-6}$&$(0.5-2.4)\times 10^{-3}$&$\lesssim 1.0\times 10^{-9}$&$\lesssim  2.2 \times 10^{-4}$&$1.8-3.7$&$0.01-0.1$&$0.5-1.3$&$(0.8-2.2)\times 10^{-3}$&$50.8-51.9$\\
\hline
\end{tabular}
}
\caption{\label{tab:values} Range of parametric values resulted by fixing $\alpha_{1}$ for the case of $\theta_S=0=\beta$, $\kappa_S= \kappa_{SS} =\alpha_{2} \equiv \alpha_{1}$, with $\mathcal{N}=1=a$, $Q=M$, $T_R \simeq 10^9$ GeV and $m_{3/2} = 10$ TeV. These ranges are ACT-compliant, i.e., they correspond to the $2\sigma$ $n_s$ bounds, $0.967\lesssim n_s \lesssim 0.980$, according to the ACT Data Release 6. }
\end{table}

As we explore larger values of $\kappa \gtrsim 0.01$ and $\alpha_1 \gtrsim 10^{-3}$, SUGRA corrections become increasingly significant due to the associated larger field values. This marks the regime where the couplings $\kappa_S$ and $\kappa_{SS}$ begin to contribute. However, their impact remains limited and subdominant under our simplifying assumption $\alpha_1 = \alpha_2 = \kappa_S = \kappa_{SS}$. In this setup, for field values approaching the Planck scale, the resulting tensor-to-scalar ratio is $r \lesssim 5 \times 10^{-6}$, which lies below the sensitivity of upcoming CMB polarization experiments \cite{LiteBIRD:2022cnt, CMB-S4:2020lpa, SimonsObservatory:2018koc}.
If we relax the above constraints and allow $\kappa_S$ and $\kappa_{SS}$ to vary independently, it becomes possible to achieve a larger and potentially observable tensor-to-scalar ratio in the range $r \lesssim 0.01$, as discussed in \cite{Shafi:2010jr, Rehman:2010wm}.
 
Figure~\ref{fig:n1a1_fixk} presents the typical behavior of inflationary parameters, specifically $n_s$ and $M$, as functions of $\alpha_1$ while keeping $\kappa$ fixed. It is evident that only a narrow window of $\alpha_1$ values yields predictions for $n_s$ that fall within the $1\sigma$–$2\sigma$ observational bounds. Among the cases considered, $\kappa = 10^{-2}$ provides the widest allowed range for $\alpha_1$, namely $8 \times 10^{-6} \lesssim \alpha_1 \lesssim 4.7 \times 10^{-5}$, as shown in figure~\ref{fig:fixed_N_varying_k_nsvalpha1}.
Interestingly, values of $\kappa$ near $0.1$, often regarded as natural, remain consistent with the central value of $n_s$ when the gravitino mass is around $10$ TeV. This stands in contrast to earlier results without $R$-symmetry violating terms (i.e., $\alpha_i = 0$) using a minimal canonical K\"ahler potential, where significantly larger soft mass scales ($\gg 10$~TeV) were needed to accommodate $\kappa \sim 0.1$ \cite{Rehman:2009nq}. However, by including $R$-symmetry preserving but non-minimal terms in the K\"ahler potential, such natural $\kappa$ values become viable even for $\kappa_S \sim 0.01$ \cite{urRehman:2006hu}.
For a discussion of natural parameter values in SHI that include $R$-symmetry breaking terms in the superpotential, see ref.~\cite{Wan:2025pcf}.

As discussed earlier in section~\ref{sec:analysis}, the choice of $\mathcal{N}$ directly affects the radiative correction term in the potential via eq.~\eqref{eq:rcorrections}, and can thus offer guidance for selecting a suitable gauge group $G$. 
In figure~\ref{fig:varing_N_fixalpha1}, we illustrate the impact of varying $\mathcal{N}$ values associated with several well-studied gauge symmetries. We find that, for a fixed $\alpha_1$, increasing $\mathcal{N}$ tends to shift and narrow the allowed ranges of $\kappa$ and the tensor-to-scalar ratio $r$ toward smaller values, while still producing values of $n_s$ compatible with ACT constraints. 
Consequently, gauge groups like $U(1)_{B-L}$ and left-right symmetry allow for broader and higher ranges of $\kappa$ and $r$, whereas flipped $SU(5)$ and $SO(10)$ produce $n_s$–$\kappa$ curves that fit observational data but correspond to unobservably small values of $r$. We now discuss the possibility of realizing the large-$r$ solution.
\section{Large-$\mathbf{r}$ Solutions and Observable Gravitational Waves} \label{sec:larger}
In this section, we relax the earlier assumption $\kappa_S = \kappa_{SS} = \alpha_2 = \alpha_1$ and explore the effect of treating the $R$-symmetry breaking parameters $\alpha_{1,2}$ and the $R$-symmetric parameters $\kappa_S$ and $\kappa_{SS}$ as independent. Our primary focus here is on identifying parameter configurations that yield large values of the tensor-to-scalar ratio $r$, potentially within the reach of upcoming gravitational wave detectors. Guided by previous and recent studies \cite{Rehman:2010wm, Rehman:2012gd, Rehman:2017gkm, Masoud:2021prr, Ahmed:2025crx}, we find that large-$r$ solutions typically arise when the inflationary potential contains a positive quadratic term and a small or negative quartic term. This behavior cannot be achieved when $|\kappa_S|$ and $\kappa_{SS}$ are small compared to $\alpha_{1,2}$. Therefore, we consider the regime where the $R$-symmetric couplings dominate over the $R$-violating ones. For simplicity, we set $\alpha_1 = \alpha_2 = \alpha$.

\begin{figure} [t!]
\centering
\subfloat[\label{fig:large_r_fixed_alpha1_rvk}]
{{\includegraphics[width=0.47\textwidth]{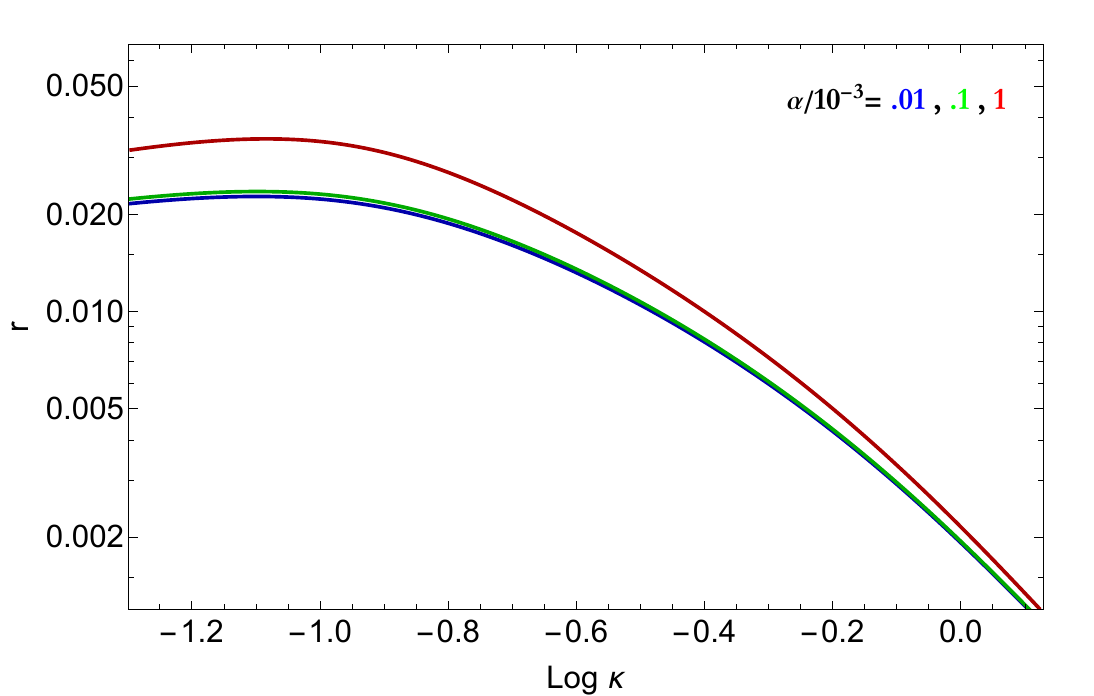}}}
\quad
\subfloat[\label{fig:large_r_fixed_alpha1_rvM}]
{{\includegraphics[width=0.47\textwidth]{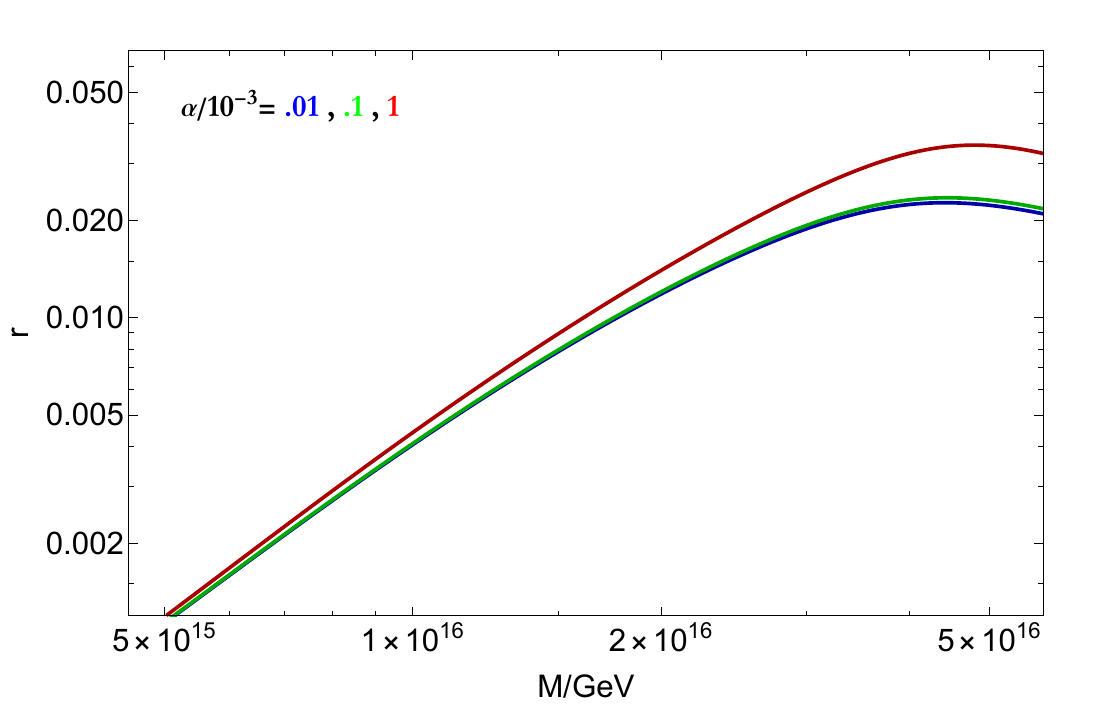}}} \\
\subfloat[\label{fig:large_r_fixed_alpha1_rvks}]
{{\includegraphics[width=0.47\textwidth]{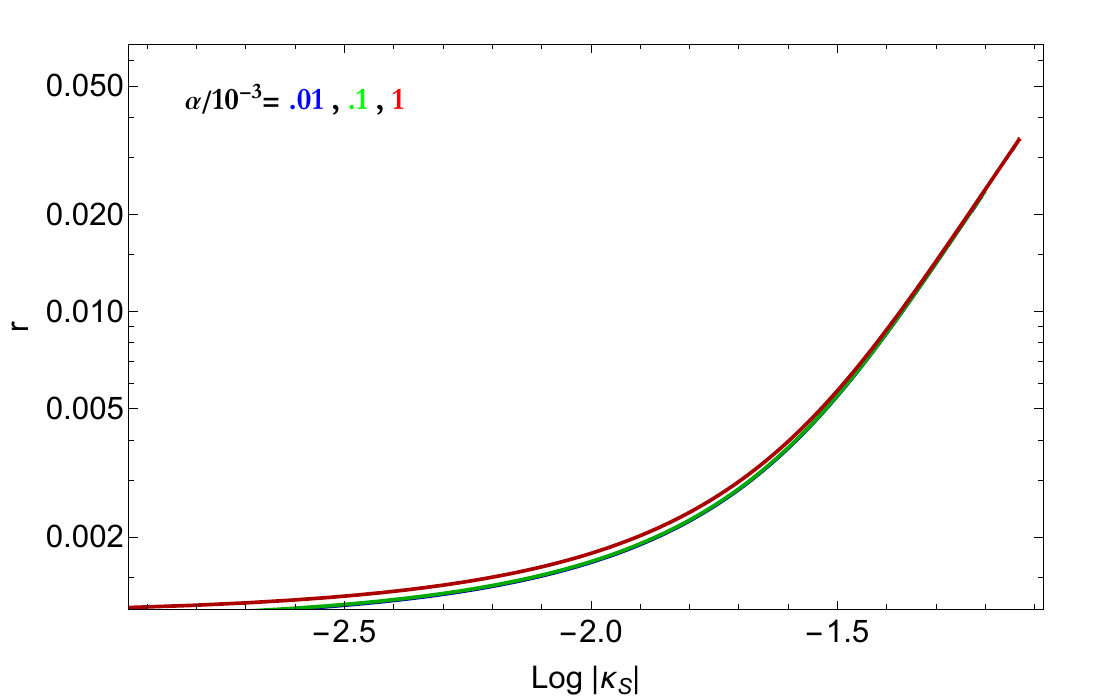}}}
\quad
\subfloat[\label{fig:large_r_fixed_alpha1_rvkss}]
{{\includegraphics[width=0.47\textwidth]{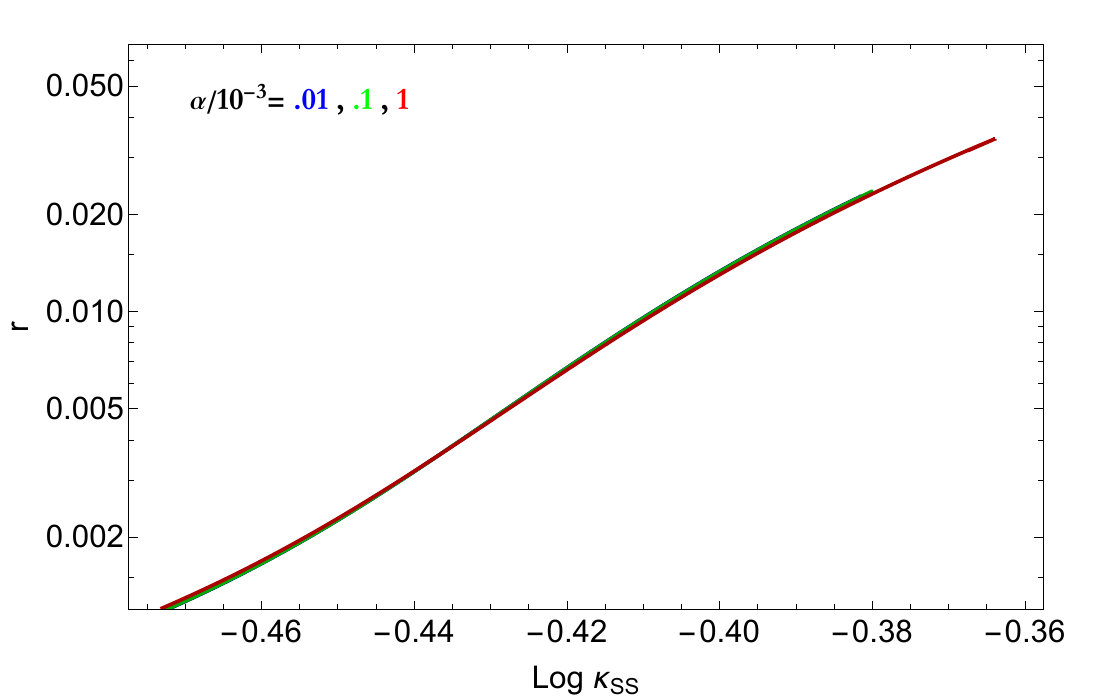}}}
\caption{Predictions for large-$r$ values assuming fixed $\alpha_1 = \alpha_2 \equiv \alpha$. These results are consistent with $n_s = 0.974$ (ACT central value), $\mathcal{N} = 1$, $a = 1$, $T_R \simeq 10^9$ GeV, $m_{3/2} = 10$ TeV, and $s_0 = m_P$. From top left to bottom right, $r$ is plotted against $\kappa$, symmetry breaking scale $M$, and the K\"ahler couplings $\kappa_S$ and $\kappa_{SS}$. Although larger $\alpha$ promotes bigger values of $r$, this effect becomes significant only when combined with non-minimal K\"ahler terms. In particular, negative values of $\kappa_S$ enhance $r$ most effectively.}
\label{fig:large_r}
\end{figure}

To probe the maximal range of $r$, we set the inflaton field value at horizon exit to $s_0 = Mx_0= m_P$. We take $\mathcal{N} = 1$, $a = 1$, $m_{3/2} = 10$~TeV, and $T_R \simeq 10^9$~GeV, and vary the parameters to extract viable large-$r$ predictions, as shown in figure~\ref{fig:large_r}. We find that for $\alpha \lesssim 10^{-5}$, the $R$-violating couplings have negligible impact on large-$r$ solutions. To demonstrate their role, we examine three representative values: $\alpha = 10^{-5}, 10^{-4}, 10^{-3}$. For even larger $\alpha$, satisfying all inflationary constraints requires delicate cancellations in the potential.
As previously noted in ref.~\cite{Rehman:2010wm}, large-$r$ solutions emerge when the quartic term becomes suppressed. This leads to a relation among the $R$-symmetric couplings:
\begin{equation}
\kappa_{SS} \simeq \frac{1}{3} \left(1 + \frac{7}{2} |\kappa_S| \right).
\end{equation}
Our findings are consistent with this, yielding $|\kappa_S| \sim 0.01 - 0.1$ and $\kappa_{SS} \sim 0.4$ in the viable region.

An important observation is that large-$r$ solutions allow for a wider and higher range of symmetry breaking scales, with $5 \times 10^{15}$~GeV $\lesssim M \lesssim 5 \times 10^{16}$~GeV. This includes the canonical GUT scale of $2 \times 10^{16}$~GeV, which was not accessible in the small-$r$ scenarios discussed earlier. In addition, the running of the scalar spectral index remains within acceptable limits, with $10^{-6} <|\alpha_s| \lesssim 10^{-2}$, consistent with the Planck observations. The predicted values of $r$ lie within the observable range $10^{-3} \lesssim r \lesssim 0.03$, making them accessible to forthcoming CMB polarization experiments such as LiteBIRD \cite{LiteBIRD:2022cnt, delaHoz:2025uae}, CMB-S4 \cite{Abazajian:2019eic, CMB-S4:2020lpa}, and the Simons Observatory \cite{SimonsObservatory:2018koc,SimonsObservatory:2025wwn}.

\section{Conclusion}\label{sec:conclusion}
In this work, we developed an SHI scenario that incorporates the effects of explicit $R$-symmetry violation on inflationary predictions, with a particular focus on nonrenormalizable terms arising in the K\"ahler potential. While earlier studies have explored such $R$-violating contributions primarily through the superpotential, our analysis extends this framework by considering an alternative setup in which the dominant $\slashed{R}$ contributions stem from the K\"ahler sector. This is realized by assuming $R$-symmetry is spontaneously broken in a hidden sector and transmitted via gravity mediation, effectively suppressing the $\slashed{R}$ terms from the superpotential at leading order.

To ensure a consistent and viable inflationary trajectory, we incorporated SUGRA corrections, soft SUSY-breaking effects, and one-loop radiative corrections. These ingredients, standard in realistic SUSY inflation models, collectively provide the necessary slope for the slow-roll dynamics of the inflaton. Our construction relies on a general gauge symmetry $G$, which we kept unspecified to maintain broad applicability, though we explored the influence of specific choices through the parameter $\mathcal{N}$ in radiative corrections. A more detailed study of particular gauge groups and their cosmological implications is left for future work.

In the numerical analysis, we simplified the potential by stabilizing the inflaton phase $\theta_S$ and assuming small, common values for the higher-order K\"ahler couplings $\kappa_S$, $\kappa_{SS}$, and $\alpha_2$. We imposed constraints from the gravitino overproduction bound by fixing $T_R \simeq 10^9$~GeV and choosing $m_{3/2} \simeq 10$~TeV, which also keeps the MSSM spectrum within the reach of collider experiments such as the LHC and FCC. The model's predictions were compared against the recent ACT Data Release 6, with a particular focus on the spectral index $n_s$ and the tensor-to-scalar ratio $r$.

Our findings show that $R$-symmetry violating terms in the K\"ahler potential, even in the absence of significant $\slashed{R}$ contributions from the superpotential, can play a central role in shaping the inflationary dynamics. The scalar spectral index $n_s$ can be brought into excellent agreement with the ACT bounds over a broad range of $\alpha_1$ and $\kappa$ values. However, the tensor-to-scalar ratio $r$ remains highly suppressed in the minimal scenario, with predictions well below $10^{-5}$—a value unlikely to be probed by current or near-future experiments.

We also examined the impact of different values of $\mathcal{N}$, associated with various gauge groups. Models with larger $\mathcal{N}$, such as those corresponding to flipped $SU(5)$ and $SO(10)$, tend to suppress both $\kappa$ and $r$, making the tensor modes even less accessible. Nonetheless, we note that by relaxing the assumption of equal K\"ahler couplings and allowing $\kappa_S$ and $\kappa_{SS}$ to vary independently, one can obtain a higher tensor-to-scalar ratio in the range $r \lesssim 0.03$, potentially within reach of future CMB observations, as previously explored in \cite{Shafi:2010jr, Rehman:2010wm}.

In short, this study highlights a viable and theoretically motivated direction for embedding controlled $R$-symmetry violation in SHI through the K\"ahler potential, offering new flexibility in matching inflationary observables while staying consistent with low-energy phenomenology.





\bibliographystyle{JHEP}
\bibliography{Bibliography}
\end{document}